\documentclass[%
 reprint,
 amsmath,amssymb,
 aps,
]{revtex4-2}

\usepackage{graphicx}
\usepackage{dcolumn}
\usepackage{bm}
\usepackage{epstopdf}
\usepackage{amsmath}
\usepackage{xcolor}
\usepackage{hyperref}

\begin{document}

\preprint{APS/123-QED}

\title{Near-extremal Kerr-like ECO in the Kerr/CFT Correspondence in Higher Spin Perturbations}
\author{M. Zhahir Djogama$^{1}$}
\email{zhahirdjo@gmail.com}
\author{Muhammad F. A. R. Sakti$^{3}$}
\email{fitrahalfian@gmail.com}
\author{Freddy P. Zen$^{1,2}$}%
\email{fpzen@fi.itb.ac.id}
\author{Mirza Satriawan$^4$}
\email{mirza@ugm.ac.id}
\affiliation{$ ^1 $Theoretical High Energy Physics Group, Department of Physics, Institut Teknologi Bandung, Bandung 40132, Indonesia}
\affiliation{$ ^2 $Indonesia Center for Theoretical and Mathematical Physics (ICTMP), Institut Teknologi Bandung, Bandung 40132, Indonesia}
\affiliation{$ ^3 $High Energy Physics Theory Group, Department of Physics, Faculty of Science, Chulalongkorn University, Bangkok 10330, Thailand}
\affiliation{$ ^4 $Department of Physics, Faculty of Mathematics and Natural Sciences, Universitas Gadjah Mada, Yogyakarta 55281, Indonesia.}

%
%
%

\date{\today}

\begin{abstract}
The Kerr/CFT correspondence has been established to explore the quantum theory of gravity in the near-horizon geometry of a extreme Kerr black holes. The quantum gravitational corrections on the near-horizon region may manifest in form of a partially reflective membrane that replace the horizon. In such modification, the black holes now can be seen as a horizonless exotic compact object (ECO). In this paper, we consider the properties of Kerr-like ECOs in near-extremal condition using Kerr/CFT correspondence. We study the quasinormal modes and absorption cross-section in that background and compare these by using CFT dual computation. The corresponding dual CFT one needs to incorporate finite size/finite $N$ effects in the dual CFT terminology. We also extend the dual CFT analysis for higher spin perturbations such as photon and graviton. We find consistency between properties of the ECOs from gravity sides and from CFT sides. The quasinormal mode spectrum is in line with non-extreme case, where the differences are in the length of the circle, on which the dual CFT lives, and phase shift of the incoming perturbation. The absorption cross-section has oscillatory feature that start to disappear near extremal limit. The particle spin determines the phase shift and conformal weight. We also obtain that the echo time-delay depends on the position of the membrane and extremality of the ECOs.
\end{abstract}

\maketitle


\section{Introduction}
The Anti-de Sitter/conformal field theory (AdS/CFT) correspondence is a significant development in string theory which is based on the idea of the holographic principle, proposed by 't Hooft \cite{tHooft}, which suggests that a higher-dimensional theory can be described by a corresponding lower-dimensional one. The AdS/CFT correspondence provides a fruitful tool for studying strongly coupled theories by relating them to weakly coupled theories and vice versa. It has been particularly successful in investigating the thermodynamics of black holes. A remarkable development of this correspondence is the successful study of the microscopic origin of Kerr black hole's entropy. It is then well known as Kerr/CFT correspondence \cite{Guica2009}. For extremal Kerr black holes, it is found that its near-horizon geometry exhibits an exact $SL(2,R)\times U(1)$ symmetry leading to the precise computation of Bekenstein-Hawking entropy from two-dimensional (2D) Cardy formula. For non-extremal Kerr black hole, the conformal symmetry does not emerge directly in the geometry of the black holes. Instead, the conformal symmetry is hidden in the solution space of a probe scalar field in the near-horizon region within the low-frequency limit approximation \cite{Castro2010}. One can read off the conformal symmetry from the quadratic Casimir operator that satisfies $SL(2,R)\times SL(2,R)$ algebra. However, the conformal symmetry is globally obstructed by the periodic identification of the azimuthal angle $\phi $. The periodic identification of $\phi$ denotes that  $SL(2,R)\times SL(2,R)$ symmetry breaks into $U(1)\times U(1)$ symmetry. The spontaneous breaking of the symmetry is caused by the left- and right-moving temperatures $T_{L,R}$. By assuming the smooth connection with the extremal black hole, the associated central charges of this non-extremal Kerr black hole are computed and used to compute the Cardy entropy. The primary result is then this Cardy entropy precisely matches with the Bekenstein-Hawking entropy of the extremal Kerr black holes.

The Kerr/CFT correspondence has come as an alternative to study the properties of classical rotating black holes. But then one can may expect that this correspondence can be applied for quantum black holes. One of the studies that considers quantum black holes in the relation with Kerr/CFT correspondence is performed in Ref. \cite{DeyAfshordiPRD2020}. Without losing the generality, we may assume that due to the strong gravitational attraction inside the black holes, quantum effect can affect the structure of the black holes. As an intriguing example provided in Ref. \cite{Oshita2020,Wang2020},  the quantum effect appears to modify the apparent horizon of the classical black holes that leads to potentially observable consequences such as the quasi-normal modes (QNMs). The structure of the event horizon is assumed to alter significantly to be infinitesimal quantum membrane. This quantum membrane is reflective and located slightly in front of the would-be horizon of the classical black hole in the order of Planck length. This novel representation is another alternative way to solve the black hole information paradox. Beside this alternative representation to information paradox, one may read some astrophysical objects that also come to solve the same problem which are fuzzballs \cite{MathurForsch2005,MathurTurtonJHEP2014}, 2-2 holes \cite{22holes}, gravastars \cite{MazurMottola2004,MazurMottolaCQG2015,SaktiSulaksonoPRD2021}, and Kerr-like wormholes \cite{BuenoCanoPRD2018}. This object is known as Kerr-like exotic compact object (ECO). The precise origin of the quantum reflective membrane is still not well grasped due to the lack of an exact calculation within a theory of quantum gravity. However, the analysis of dual CFT on this ECO is portrayed in terms of Boltzmann factor that matches with the reflectivity of the quantum horizon \cite{OshitaPRD2020}.

The existence of the quantum reflective membrane gives rise to some potential observables. The principal potential observable from this feature is the presence of gravitational echoes. The gravitational echoes may be detected in the postmerger ringdown signal of a binary system coalescence such as black hole coalescence \cite{CardosoFranzinPRL2016,OshitaAfshordiPRD2019,CardosoHopperPRD2016,AbediPRD2017,Wang2020}. In particular, the major realizations for such observable detection of echoes lies in the detection of QNMs since the ringdown phase is dominated by this modes. The echo signals that bring these modes are separated from the primary ringdown signal by the corresponding time-delay due to the presence of the reflective membrane. In several papers, it is claimed that the potential evidence of gravitational echoes has been discovered within LIGO data \cite{WangAfshordiPRD2018echology,ConklinHoldomPRD2019,AbediAfshordiJCAP2019,ConklinHoldomRenPRD2018}.

It is pointed out in Ref.~\cite{AbediPRD2017} that the non-linear physics may affect the time between the main merger event and the first echo. Due to non-linear physics effect, the magnitude of the time-delay may be shifted around 2$\%$-3$\%$. As an example, this shift can occur in Rastall theory of gravity where the covariant derivative of the matter tensor is not null and proportional to an arbitrary constant that corresponds with the time-delay shift \cite{SaktiSurosoSulaksonoZen2022}. Not only the time-delay, the analysis of QNMs in the ringdown spectrum can be a proper procedure to verify the natural structure of black holes or ECOs. Furthermore, this also can be another fascinating playground to probe some features beyond general relativity.
The calculation of QNMs might probe the extra dimension from the ECOs in such braneworld scenario \cite{DeyChakrabortyPRD2020,DeyBiswasPRD2021}. 

In near future, a number of experiments to improve the precision related to the study of astrophysical objects and phenomena will run such as will be done by LIGO or Event Horizon Telescope. More specifically, they will run some experiments to advance black hole's observations and related to its properties. Kerr (rotating) black hole is considered as the most physical black hole in the universe. A huge number of observations denotes that the observed black holes are rotating very fast, nearly the extremal limit, especially the supermassive ones in the active galactic nuclei \cite{ReynoldsSpaceSciRev2014,BrennemanSpringer2013}. As a very specific example, the near-extremal Kerr is found to be the source of X-ray in the binary system GRS 1915+105 \cite{McClintocketalAstropJ2006}. Therefore, it is very relevant to study the near-extremal black hole and its quantum counterpart (ECOs).

In this work, we will consider near-extremal Kerr ECO. We will study the scalar field in that background. Instead of imposing purely ingoing boundary condition, we impose reflective boundary conditions at the place slightly near the event horizon of the near-extremal Kerr metric (would-be horizon). Due to the scalar field perturbation, we compute the QNMs by solving the propagation of massless scalar field modes. We then compare the results by using CFT dual computation as has been done for generic Kerr ECO. Moreover, we extend the calculation for higher spin fields such as photon and graviton. We want to see the echoes emerging in the absorption cross-section of those fields. In the end, we will compute the echo time-delay produced by the propagation of the fields on the near-extremal Kerr ECO background.

The organization of this paper is given as follows. In the section \ref{sec2}, we review ECO in the near-extremal Kerr ECO background and assume the reflective boundary condition to compute the absorption cross-section and QNMs. We also review the current understanding of hidden conformal symmetry of classical Kerr. Then in Section \ref{sec5}, we construct dual CFT of ECO and compare the result from gravity computation with the CFT result. In Section \ref{sec6}, we perform similar computation for higher spin fields including their QNMs. Then in Section \ref{sec7}, the echo time-delay is computed for all fields. Finally, we summarize our work in the last section.

\section{Kerr-like Exotic Compact Object}\label{sec2}

We consider ECO with Kerr metric as its exterior spacetime. The near-horizon geometry is modified by replacing the event horizon with partially reflective membrane as a consequence of a quantum gravitational effect \cite{Oshita2020,Wang2020}. This membrane is located slightly outside the usual position of the would-be horizon $(r_++\delta r)$, where $\delta r$ is to be the order of Planck length for stability of this ECO \cite{Maggio2017,Maggio2019}. For a rotating Kerr-like ECO with mass $M$ and angular momentum $J=aM$, the metric in the Boyer-Lindquist coordinate can be written as
\begin{align}\label{kerrmetric}
	ds^2 = &-\left(1-\frac{2M\hat{r}}{\hat{\rho}^2}\right)d\hat{t}^2 \nonumber\\
	&+ \left(\hat{r}^2+a^2+\frac{2a^2M\hat{r}\sin^2\theta}{\hat{\rho}^2}\right)\sin^2\theta d\hat{\phi}^2 \nonumber\\ 
	&-\frac{4aM\hat{r}\sin^2\theta}{\hat{\rho}^2}d\hat{\phi} d\hat{t} + \frac{\hat{\rho}^2}{\Delta}d\hat{r}^2 +\rho^2d\theta^2,
\end{align}
where
\begin{equation}
	\hat{\rho}^2 = \hat{r}^2 + a^2\cos^2\theta,~~~
	\Delta = \hat{r}^2+a^2-2M\hat{r}.
\end{equation}
The usual position of the horizons and the angular velocity on the horizon are given by
\begin{equation}
r_\pm = M \pm \sqrt{M^2-a^2}, ~~\Omega_H = \frac{a}{2Mr_+}.\
\end{equation}
The Hawking temperature and the Bekenstein-Hawking entropy are given by
\begin{equation}
T_H = \frac{r_+-r_-}{8\pi Mr_+},~~S_{BH} =2\pi Mr_+.\
\end{equation}

Two models of the membrane are studied recently. These are the constant reflectivity and Boltzmann frequency-dependent reflectivity \cite{Oshita2020,Wang2020}
\begin{equation}\label{reflectivity}
	\mathcal{R} = \begin{cases}
		R_c, & \mbox{constant reflectivity}\\
		e^{-\frac{|\omega-m\Omega_H|}{2T_{QH}}}, & \mbox{Boltzmann reflectivity}\\
	\end{cases}
\end{equation}
where $R_c$ is a constant and $T_{QH}$ is defined as "quantum horizon temperature" \cite{OshitaPRD2020} and expected to be comparable to the Hawking temperature with an arbitrary proportional $\gamma$ $(T_{QH}=\gamma T_H)$  \cite{Oshita2019}. We will see how this modification of boundary condition at the horizon affects the absorption cross-section and  QNM spectrum of Kerr-like ECO in near-extremal condition.

\subsection{Near-extremal Kerr ECO Background}
\label{sec3}
One of the realization of the correspondence between Kerr black holes and the CFTs is the equivalence of the absorption cross-section \cite{Castro2010}. It turns out that this also true for Kerr-like ECO as shown in \cite{DeyAfshordiPRD2020}. They show that the absorption cross-section for Kerr-like ECO has an additional factor compared to that of black holes, related to the reflectivity, and this part can also emerge from 2D CFT living on a finite circle. Another quantity that we can investigate is QNM, where the reflectivity of the ECO affects the imaginary part of QNM. As both quantities depend on the reflectivity, we cannot directly take extremal condition to explore the scattering issue in general because the Boltzmann reflectivity will become zero. However, for near-extremal case, we may explore this scattering problem. So in this section, we focus on near-extremal limit of the Kerr-like ECO.

To study the conformal symmetry, we consider scattering of a massless scalar field expanded in modes
\begin{equation}
	\Phi = e^{im\hat{\phi}-i\omega\hat{t}}S(\theta)R(\hat{r}).
\end{equation}
The wave equations for a full Kerr metric (\ref{kerrmetric}) are
\begin{equation}
	\frac{1}{\sin\theta}\partial_\theta(\sin\theta\partial_\theta S)+\left(K_l-\frac{m^2}{\sin^2\theta}-a^2\omega^2\sin^2\theta\right) S = 0,
\end{equation}
and
\begin{equation}\label{radialeqkerr}
	\partial_{\hat{r}}(\Delta\partial_{\hat{r}} R)+\left(\frac{(\omega(\hat{r}^2+a^2)-ma)^2}{\Delta}+2ma\omega-K_l\right)R=0,
\end{equation}
where $K_l$ is separation constant. Define the dimensionless coordinate $x$ and dimensionless Hawking temperature $\tau_H$ \cite{Bredberg2010}
\begin{align}\label{nearexcoord}
	x &= \frac{\hat{r}-r_+}{r_+}, & \tau_H = \frac{r_+-r_-}{r_+}=8\pi MT_H,
\end{align}
where the near-extremal regime corresponds to $\tau_H\ll1$. In this regime, the radial equation (\ref{radialeqkerr}) becomes
\begin{equation}\label{radialeqkerr2}
	x(x+\tau_H)R''+(2x+\tau_H)R'+V_lR=0,
\end{equation}
where the prime denotes $\partial_x$ and
\begin{equation}
	V_l=\frac{\left[x(x+2)m+\tau_H\tilde{\omega}\right]^2}{4x(x+\tau_H)}+m^2-K_l,
\end{equation}
\begin{equation}
	\tilde{\omega}=\frac{\omega-m\Omega_H}{2\pi T_H}.
\end{equation}
In the near-extremal limit, $\tilde{\omega}$ is held fixed when $T_H\to0$. This means only modes with energy near superradiant bound $\omega\simeq\Omega_H$ will survive. Modes with energies outside the scale of the bound will not cross into the near-horizon region. In the Appendix \ref{app:scalwave}, the radial equation is solved in far region $x\gg\tau_H$ and near region $x\ll1$, then matched in the matching region $\tau_H\ll x\ll1$. For far and near regions, the solution is similar to the near-extremal black holes case \cite{Bredberg2010, Hartman2010} as the exterior background of ECO is still the classical Kerr. The difference becomes apparent when we impose new boundary condition (\ref{membound}) at the matching region to include the reflective membrane.

\subsection{Absorption cross-section}
The absorption cross-section is the ratio between absorbed flux at the event horizon and incoming flux from infinity
\begin{equation}
	\sigma_{abs}=\frac{\mathcal{F}^{abs}_{r\to r_+}}{\mathcal{F}^{in}_{\infty}}\sim \tilde{\omega}\tau_H\frac{1-|\mathcal{R}|^2}{\left|\frac{A}{C}+\frac{B}{C}\right|^2}.
\end{equation}
In this calculation we only consider near-horizon contribution and case of real $\beta$ to match with CFT (far-region contribution and case of imaginary $\beta$ is discussed in \cite{Bredberg2010}). Since we consider near-extremal limit $\tau_H\ll1$, $|A/C|$ is more dominant than $|B/C|$. Using $|A/C|$ given by (\ref{ACmatching}), we find
\begin{align}\label{absgrav}
	\sigma_{abs} &\sim \frac{(\tau_H)^{2\beta}}{\pi\Gamma(2\beta)^2}\sinh\left(\pi \tilde{\omega}\right)\nonumber\\
	&\times\left|\Gamma\left(\frac{1}{2}+\beta+im\right)\Gamma\left(\frac{1}{2}+\beta+i(\tilde{\omega}-m)\right)\right|^2\nonumber\\
	&\times\frac{1-|\mathcal{R}|^2}{\left|1-\mathcal{R}e^{-2ir_0^*(\omega-m\Omega_H)+i\pi\delta}\right|^2},
\end{align}
where $r_0^*=r^*(\hat{r}_0)$ is the position of membrane in tortoise coordinate defined as
\begin{equation}
	r^* = \int \frac{\hat{r}^2+a^2}{(\hat{r}-r_+)(\hat{r}-r_-)}dr,
\end{equation}
which then we get 
\begin{equation}\label{r0}
	\ln\left(\frac{x_0}{\tau_H}\right)=\ln\left(\frac{\delta r}{r_+-r_-}\right)\sim r_0^*\frac{(r_+-r_-)}{(r_+^2+a^2)}.
\end{equation}
The difference between absorption cross-section obtained for ECOs and black holes is the oscillatory feature, in addition to the classical cross-section, that depends on $\mathcal{R}$. Obviously for $\mathcal{R}=0$, it is associated with the classical black hole. 

\subsection{Quasinormal modes}
The oscillation of radiation from ECO corresponds to the resonance at the ECO's QNM frequencies. The parameters contained by the ECOs affect the QNM spectrums. Therefore, since those parameters also affect the near-horizon quantum structure of the ECOs, QNM may contain significant information about it, as shown in Ref.~\cite{SaktiSurosoSulaksonoZen2022}. There are various methods used to obtain the QNM spectrums of black holes or ECOs. Usually, a purely outgoing boundary condition at the asymptotic infinity is imposed. However, it is difficult to solve the wave equation governing the perturbations analytically, making it impractical to get the QNMs of ECOs without making assumptions. Another method to obtain QNM spectrum, especially using AdS/CFT duality, is from poles of retarded Green's function \cite{Mark2017}. In this calculation, we use the low-frequency limit to solve approximately the wave function and obtain the QNM spectrums.

Purely outgoing boundary condition means that there is no incoming waves from infinity. Using this condition and matching (\ref{radialsolfar1}) and (\ref{radialsolnear1}) give us
\begin{equation}\label{DC}
	\frac{D}{C} = -\frac{P_-+\sigma Q_-}{P_++\sigma Q_+},
\end{equation}
where 
\begin{equation}
	\sigma = (\tau_H/im)^{-2\beta}\frac{\Gamma(2\beta)\Gamma(1+2\beta)\Gamma(1-\beta-im)}{\Gamma(-2\beta)\Gamma(1-2\beta)\Gamma(1+\beta-im)}.
\end{equation}
\textcolor{black}{Together with (\ref{membound}), near the horizon we have the following expression
\begin{equation}
	\mathcal{R}e^{i\pi\delta}=-\frac{P_-+\sigma Q_-}{P_++\sigma Q_+}x_0^{i\tilde{\omega}}.
\end{equation}
In general this equation cannot be solved directly, we need to use numerical method to obtain the QNM.} However, near the superradiant bound $\omega\simeq m\Omega_H$ we can solve it analytically in the low-frequency limit $M\omega\ll1$. In this limit we get QNM spectrums
\begin{align}\label{qnm}
	\omega-m\Omega_H \simeq &\frac{1}{2r_0^*}\pi(2n+1+\delta)\nonumber\\
	&\times\left(1-\frac{i\times\text{sgn}[2n+1+\delta]}{4r_0^*\gamma T_H}\right).
\end{align}
where $n$ is a positive integer and we consider the reflectivity as the Boltzmann reflectivity. This result has the same explicit form with the QNM spectrum of Kerr-like ECO for non-extreme case \cite{DeyAfshordiPRD2020}.

\subsection{Hidden conformal symmetry}
\label{sec4}

It is already well known that Kerr black hole is dual to 2D CFT \cite{Guica2009}. However, unlike extremal Kerr black holes, the conformal symmetries are hidden for non-extremal case \cite{Castro2010}. They are hidden locally on the radial scalar wave equation and globally broken under periodic identification of $\hat{\phi}\to\hat{\phi}+2\pi$. The symmetry becomes apparent when we investigate the scalar wave equation at the near-horizon region. For more example on extremal Kerr/CFT correspondence, one can find in \cite{Sakti2018,SaktiEPJPlus2019,SaktiMicroJPCS2019,SaktiPhysDarkU2020,SaktidyonicKerrSen} and for non-extremal one in \cite{SaktiEPJC2024,Saktideformed2019,SaktiNucPhysB2020,SaktiEPJC2023} cases.  For a review, one can find in Ref. \cite{Compere2017}.

As we see in the near region radial equation, the solution is given by the hypergeometric functions, which fall under $SL(2,R)$ representation. This give us a hint to investigate the conformal symmetries. In order to understand this hidden symmetry, we can introduce the following set of conformal coordinates \cite{Castro2010, Haco2018}
\begin{align}
	w^+ &=\sqrt{\frac{\hat{r}-r_+}{\hat{r}-r_-}}e^{2\pi T_R\hat{\phi}},\\
	w^- &=\sqrt{\frac{\hat{r}-r_+}{\hat{r}-r_-}}e^{2\pi T_L\hat{\phi}-\frac{\hat{t}}{2M}},\\
	y &=\sqrt{\frac{\hat{r}_+-r_-}{\hat{r}-r_-}}e^{\pi(T_R+T_L)\hat{\phi}-\frac{\hat{t}}{2M}},
\end{align} 
where we identify right- and left-moving temperatures as
\begin{align}\label{cfttemp}
	T_R&=\frac{r_+}{4\pi a}\tau_H, & T_L=\frac{r_++r_-}{4\pi a}.
\end{align}
The right-moving temperature is proportional to near-extremal Hawking temperature. In terms of these new coordinates, we can define two sets of vector fields to form a quadratic Casimir operator. It is shown in Ref.~\cite{Castro2010} that in terms of $(\hat{t},\hat{r},\hat{\phi})$, we can express the scalar wave equation at the near-horizon region for Kerr background (\ref{kerrmetric}) as that $SL(2,R)$ Casimir operator.

For fixed $r$, we can write the relation between the conformal coordinates and Boyer-Lindquist coordinates as
\begin{equation}
	w^\pm=e^{\pm t_{R,L}},
\end{equation}
where we define
\begin{align}\label{torusco}
	t_R &= 2\pi T_R\hat{\phi}, & t_L=\frac{\hat{t}}{2M}-2\pi T_L\hat{\phi}.
\end{align}
This relation is analogous to the relation between Minkowski coordinates $(w^\pm)$ and Rindler coordinates $(t_{R,L})$. The frequencies $(\omega_L,\omega_R)$ related with the Killing vectors $(i\partial_{t_L},i\partial_{t_R})$ are conjugate to $(t_L,t_R)$.

In order to check the consistency between absorption cross-section or QNMs from gravity and CFT, we need to choose left and right frequencies $\tilde{\omega}_L$, $\tilde{\omega}_R$. One way of writing the relation between these frequencies and $(\omega,m)$ of Kerr spacetime is by considering the first law of black hole's thermodynamics $T_H\delta S_{BH}=\delta M-\Omega_H\delta J.$ We identify $\delta M$ as $\omega$ and $\delta J$ as $m$, and consider conjugate charges $(\delta E_L,\delta E_R)$,
\begin{equation}
	\delta S_{BH}=\frac{\delta E_R}{T_R}+\frac{\delta E_L}{T_L},
\end{equation} 
which leads to identification of $\delta E_{L,R}$ as $\bar{\omega}_{L,R}$. Thus, relations between $(\omega,m)$ and $\delta E_{L,R}$ are
\begin{align}\label{cftfrequency}
	\delta E_L &= \tilde{\omega}_L = 2\pi T_L\omega_L = \frac{2M^2}{a}\omega,\nonumber\\
	\delta E_R &= \tilde{\omega}_R = 2\pi T_R\omega_R = \tilde{\omega}_L-m.
\end{align}

\section{Dual CFT of Kerr-like ECO}
\label{sec5}

\subsection{From BH to ECO}
Previously we reviewed the current understanding of hidden conformal symmetry of classical Kerr black holes. Now we will construct the dual CFT for Kerr-like ECO following procedure in Ref. \cite{DeyAfshordiPRD2020}. For ECOs, the near-horizon geometry is modified due to quantum gravitational effects. The horizon is replaced by partially reflective membrane located slightly outside the would-be horizon. Although the near-horizon region is modified, the exterior spacetime is identical to classical Kerr. Thus in ECOs case, the solution of scalar perturbation is the same with classical one which the solution is written in hypergeometric function, hinting the hidden conformal symmetry as well. As for the horizon modification, it can be seen as finite size/finite $N$ effects in AdS/CFT perspective \cite{Birmingham2003,Solodukhin2005,Kabat2014}. Based on this effects, the CFT now live on a space with finite volume which should affects the QNM and absorption cross-section. 

To obtain QNM spectrums and later absorption cross-section, we start with two-point function of a CFT living on a torus with spatial cycle of length $L$ and temporal cycle of length $1/T$. Unlike the usual description Kerr/CFT correspondence where cylindrical approximation of torus $(L\gg1/T)$ is used, in this case to accommodate the finite size/finite $N$ effects we keep both spatial and temporal coordinates to be finite. Hence, we need doubly-periodic two-point function on a torus to obtain the physical quantities. 

Modification of the size of space where CFT lives is done by adding an integer factor $L$ to the periodicity of azimuthal angle in the torus coordinates. Therefore, the CFT is put on a circle of finite size. Moreover, the spatial length factor $L$ corresponds to the length of that circle. We then also apply additional thermal periodicity of imaginary time to the azimuthal periodicity,
\begin{align}
	\hat{\phi} &\to \hat{\phi}+2\pi L+i\frac{\Omega_H}{T_H}, & \hat{t}=\hat{t}+\frac{i}{T_H}.
\end{align}
With this identification, the periodicity of CFT coordinates (\ref{torusco}) becomes
\begin{equation}
	t_L \to t_L + -4\pi^2LT_L-i\left(\frac{2\pi T_L\Omega_H}{T_H}-\frac{1}{2MT_H}\right),\nonumber 
\end{equation}
\begin{equation}\label{torusperiod}
	t_R \to t_R + 4\pi^2LT_R+i\frac{2\pi T_R\Omega_H}{T_H}.
\end{equation}

Since our computation here based on the torus coordinate and its periodicity (\ref{torusperiod}), we need to derive two-point function on the torus. The general form of two-point function on the torus is described in Appendix \ref{app:2pttorus}. Different with two-point function in the usual Kerr black holes, the two-point function in this case is derived through the method of images. By using this method, the Fourier transformation of the CFT two-point function is given as
\begin{align}\label{2ptf1}
	\tilde{G}(\omega_L,\omega_R)&=\int dt_Ldt_R e^{-i\omega_Lt_L}e^{-i\omega_Rt_R}\nonumber\\
	&\times\sum_{p\in\mathbb{Z}}\frac{(\pi T_L)^{2h_L}}{\left[\sinh\pi T_L\left(\frac{t_L}{2\pi T_L}+p(2\pi L+\frac{i}{T_L})\right)\right]^{2h_L} },\nonumber\\ 
	&\times\frac{(\pi T_R)^{2h_R}}{\left[\sinh\pi T_R\left(\frac{t_R}{2\pi T_R}+p(2\pi L+\frac{i}{T_R})\right)\right]^{2h_R}},
\end{align}
The infinite sum appears in the two-point function because of the method of images, as we are shifting the periodicity by an integer multiples and summing over the images to take into account the correct periodic identification (\ref{torusperiod}) of the torus. Moreover, this CFT two-point function matches with the general two-point function on the torus (\ref{2pttorus}) if we choose the modular parameters as ($\textbf{a}=1, \textbf{b}=1, \textbf{c}=-1, \textbf{d}=0$). However, later when we compute the QNMs, we will see that we can generalize the two-point function by not fixing the value of modular parameter $\textbf{a}$. For the convenience, we define new torus coordinate $\tilde{t}_{R,L}$ as
\begin{equation}
	\tilde{t}_{L,R} = \frac{t_{L,R}}{2\pi T_{L,R}}+p(2\pi L+i/T_{L,R}).
\end{equation}
Thus the two-point function becomes
\begin{eqnarray}\label{2ptftorus}
	\tilde{G}(\tilde{\omega}_L,\tilde{\omega}_R) &=& \sum_{p\in\mathbb{Z}} e^{ip(2\pi L(\tilde{\omega}_L+\tilde{\omega}_R)+i\left(\frac{\tilde{\omega}_L}{T_L}+\frac{\tilde{\omega}_R}{T_R}\right))}\int d\tilde{t}_Ld\tilde{t}_R\nonumber\\ 
	&& \times\frac{e^{-i\tilde{\omega}_L\tilde{t}_L}e^{-i\tilde{\omega}_R\tilde{t}_R}(\pi T_L)^{2h_L}(\pi T_R)^{2h_R}}{\left[\sinh(\pi T_L\tilde{t}_L)\right]^{2h_L} \left[\sinh(\pi T_R\tilde{t}_R)\right]^{2h_R}},\nonumber\\
	&\propto& T_L^{2h_L-1}T_R^{2h_R-1}e^{-\frac{\tilde{\omega}_L}{2T_L}-\frac{\tilde{\omega}_R}{2T_R}}\nonumber\\ 
	&&\times\left|\Gamma\left(h_L+i\frac{\tilde{\omega}_L}{2\pi T_L}\right)\Gamma\left(h_R+i\frac{\tilde{\omega}_R}{2\pi T_R}\right)\right|^2\nonumber\\
	&&\times \left[\frac{1}{1-e^{i2\pi L(\tilde{\omega}_L+\tilde{\omega}_R)-\left|\frac{\tilde{\omega}_L}{T_L}+\frac{\tilde{\omega}_R}{T_R}\right|}}\right.\nonumber\\
	&&\left.-\frac{1}{1-e^{i2\pi L(\tilde{\omega}_L+\tilde{\omega}_R)+\left|\frac{\tilde{\omega}_L}{T_L}+\frac{\tilde{\omega}_R}{T_R}\right|}}\right].
\end{eqnarray}

\subsection{QNMs from CFT}
From AdS/CFT duality perspective, QNM spectrums are given by the poles of retarded CFT correlation function. By employing this duality, the QNM computation is consistent with the gravity result as shown in Refs.~\cite{SaktiNucPhysB2020,Birmingham2002, ChenChu2010, ChenLong2010}. Furthermore, discrete QNMs spectrums are believed to be produced by finite size/finite $N$ effects on the CFT side. The QNM spectrums for ECOs come from the poles of the exponential part of the CFT two-point function. There are two poles lying in the upper and lower half of the $\omega$ plane. Other poles coming from the singularities of the Gamma function correspond to the usual Kerr black hole QNM spectrums. In this case, QNM spectrums come from the retarded correlation function, so the poles of (\ref{2ptftorus}) in the lower half plane are the one that we need.
Based on the definition of CFT temperatures (\ref{cfttemp}) and CFT frequencies (\ref{cftfrequency}), taking near-extremal and near-superradiant limit, the QNM spectrum is
\begin{eqnarray}\label{qnmcft}
	\omega-m\Omega_H & = &\frac{1}{8ML}(2n-2mL)\nonumber\\
	&\times&\left(1-\frac{i\times \text{sgn}[2n-2mL]}{8M\pi LT_H}\right).
\end{eqnarray}
The QNM spectrum for near-extremal Kerr-like ECO is approximately calculated in the low-frequency limit as provided in (\ref{qnm}). If we consider $\gamma=1/2$ in the gravity side, both real and imaginary part of QNM for ECOs would match with the CFT result if we define
\begin{align}\label{toruslength}
	L &= \frac{|r_0^*|}{4\pi M}, & \delta &= -1-2mL.
\end{align}
However, this case is very specific for $\gamma=1/2$.

Since gravity/CFT duality in the finite size/finite $N$ effects is not yet well defined, we can take a guess to some extent on how this effects describe the dual CFT correspondence. As long as the result match with the gravity computation. That being the case, we can make our result more general by keeping the value of modular parameter $\textbf{a}$ in (\ref{2ptf1}) to be general. In this way, the two-point function is now modified as \cite{DeyAfshordiPRD2020}
\begin{eqnarray}\label{2ptftorus2}
	\tilde{G}(\tilde{\omega}_L,\tilde{\omega}_R)
	&\propto& T_L^{2h_L-1}T_R^{2h_R-1}e^{-\frac{\tilde{\omega}_L}{2T_L}-\frac{\tilde{\omega}_R}{2T_R}}\nonumber\\ 
	&&\times\left|\Gamma\left(h_L+i\frac{\tilde{\omega}_L}{2\pi T_L}\right)\Gamma\left(h_R+i\frac{\tilde{\omega}_R}{2\pi T_R}\right)\right|^2\nonumber\\
	&&\times \left[\frac{1}{1-e^{i2\pi L(\tilde{\omega}_L+\tilde{\omega}_R)-\textbf{a}\left|\frac{\tilde{\omega}_L}{T_L}+\frac{\tilde{\omega}_R}{T_R}\right|}}\right.\nonumber\\
	&&\left.-\frac{1}{1-e^{i2\pi L(\tilde{\omega}_L+\tilde{\omega}_R)+\textbf{a}\left|\frac{\tilde{\omega}_L}{T_L}+\frac{\tilde{\omega}_R}{T_R}\right|}}\right].
\end{eqnarray}
Therefore, the general form of QNM spectrum of Kerr-ECO from CFT is
\begin{eqnarray}\label{qnmcft2}
	\omega-m\Omega_H & = &\frac{1}{8ML}(2n-2mL)\nonumber\\
	&\times&\left(1-\frac{i\textbf{a}\times \text{sgn}[2n-2mL]}{8M\pi LT_H}\right).
\end{eqnarray}
Compared with gravity computation, we get $L$ and $\delta$ the same as before and also new relation between \textbf{a} and $\gamma$ as
\begin{equation}\label{agamma}
	\textbf{a}=\frac{1}{2\gamma}.
\end{equation}
This relation describes that the property of the membrane (in this case is $\gamma$) is represented by modular parameter \textbf{a} in the CFT side. Notice that the paramater \textbf{a} is related with scale transformation in the modular transformation, this might explain on how the presence of reflective membrane can modify the near-horizon geometry by the finite size/finite $N$ effects from AdS/CFT perspective.

Compared with non-extreme case, this QNM spectrum has different dependency of spin of the ECO, since $a$ and $M$ is almost interchangeable in this case. In fact, if we take $a\simeq M$ condition of Eq. (40) in \cite{DeyAfshordiPRD2020}, it produces the same QNM spectrum with (\ref{qnmcft2}). Because of this difference, we also have different definition of $L$ and $\delta$ which contribute to the cross-section and echo time-delay. With this identification, we can find from the imaginary part of QNM (\ref{qnmcft2}) that reflectivity of the membrane is
\begin{equation}
	\mathcal{R} = e^{-|\omega-\Omega_H|/2\gamma T_H},
\end{equation}
which match exactly with Boltzmann reflectivity.
\begin{figure*}[!t]
	\centering
	\includegraphics[scale=0.5]{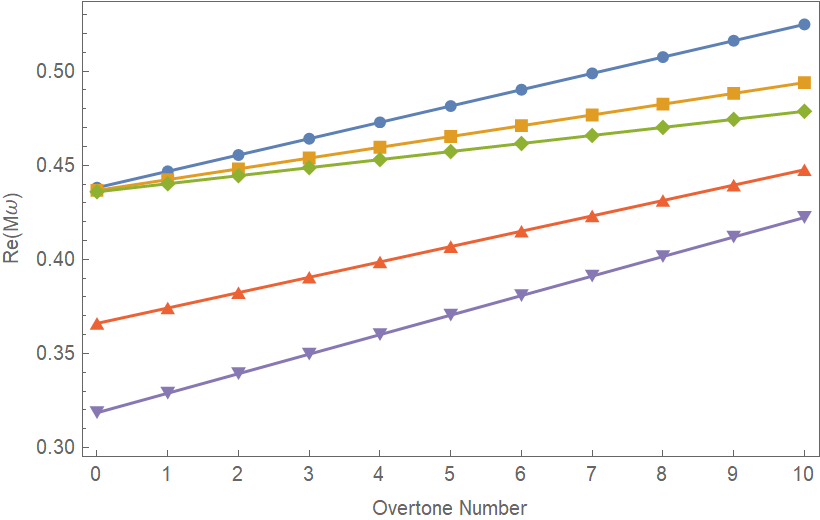}
	\hspace{1em}
	\includegraphics[scale=0.5]{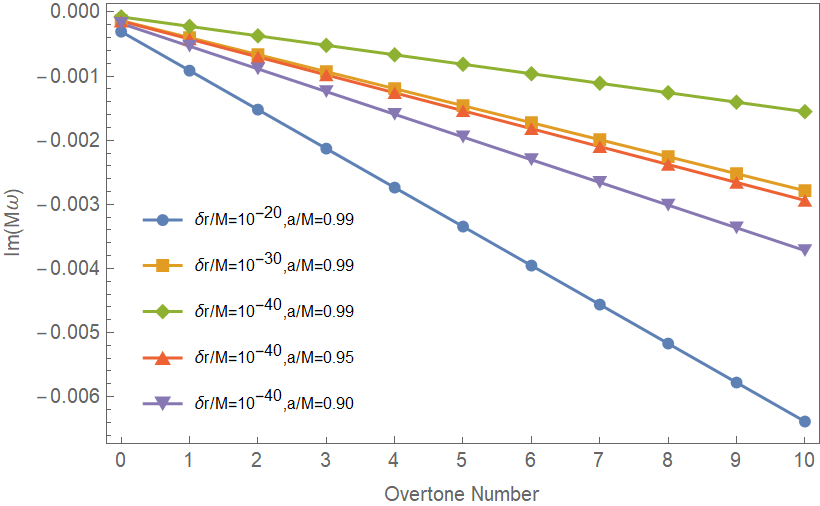}
	\caption{Real and imaginary part of QNM as a function of overtone number $n$ with $a/M$ and $\delta r/M$ variations. We set $m=1$, $\delta=0$, and $\textbf{a}=1/2$.}
	\label{fig:qnm}
\end{figure*}

\begin{figure*}[!t]
	\centering
	\includegraphics[scale=0.68]{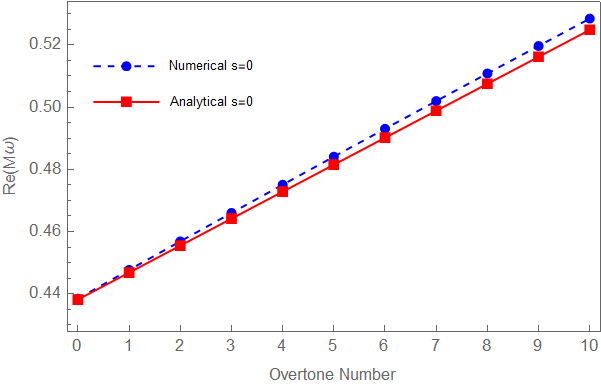}
	\hspace{1em}
	\includegraphics[scale=0.68]{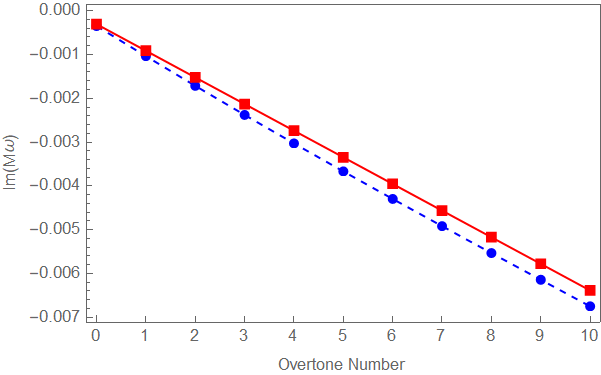}
	\textcolor{black}{\caption{Real and imaginary part of QNM computed numerically and analytically. We set $a/M=0.99$, $\delta r/M=10^{-20}$, $m=1$, $\delta=0$, and $\textbf{a}=1/2$.}
	\label{fig:qnmnum}}
\end{figure*}

QNMs that normalized by the mass $M$ depend on three parameters: the ECO spin $a/M$, the position of the membrane in tortoise coordinate $r_0^*$, and the modular paramater \textbf{a}. Furthermore according to Eq. (\ref{r0}), $r_0^*$ depends on the distance $\delta r/M$ which equivalent to the compactness $M/r_0$. Other than that, QNMs also depend on three integer number: the angular number $l$, the azimuthal number $m$, and the overtone number $n$. In Fig. (\ref{fig:qnm}), we focus on $l=m=1$ and $\textbf{a}=1/2$ to plot the QNMs to the overtone number with variation of $a/M$ and $\delta r/M$. We can see that nearer extremal limit (higher $a/M$) produces higher $\omega$ and smaller $L$ (higher $\delta r/M$) also produces higher $\omega$. The imaginary part is much smaller than the real part $Im(\omega)\ll Re(\omega)$ thus we can consider the imaginary part as perturbation from the reflectivity membrane to the QNMs. Modes with low-frequency are more dominant in late times which will lead into echoes production. \textcolor{black}{In Fig. (\ref{fig:qnmnum}), we compare the QNM result from numerical and analytical method with $a/M=0.99$ and $\delta r/M=10^{-20}$. We can see the low-frequency approximation is plausible for lower overtone modes.}

From QNMs, we can also see the condition of ergoregion instability. The instability occurs due to infinite amplification of trapped incoming waves between reflective membrane and potential barrier, inside the ergoregion. When the wave is reflected by the membrane or potential barrier and then cross through the ergoregion, it will gain energy from the ECO through Penrose's process. Since the wave is trapped, the amplification will arise exponentially, causing instability. From the point of view of the QNMs, this instability corresponds to positive imaginary part $Im(\omega)>0$. To avoid this, ECO needs to have some absorption of the incoming waves, in other words the membrane should be partially reflective \cite{Maggio2017,Maggio2019}. From Fig. (\ref{fig:qnm}), we can see in our case the instability is avoided for all modes by choosing the modular parameter $\textbf{a}=1/2$ $(\gamma=1)$ or positive in general. As for the real part, modular parameter does not affect it.
 
\subsection{Absorption cross-section from CFT}
In the previous section, we have seen the contribution of reflective membrane of the ECOs that is the emergence of new poles in the two-point function. The identification (\ref{toruslength}) makes QNM from CFT consistent with gravity calculation. In Eq. (\ref{absgrav}), the presence of the reflective membrane contributes as an oscillatory feature on the absorption cross-section. We need to check if the identification (\ref{toruslength}) can produce the same near-horizon contribution to the absorption cross-section on CFT computation. 

\begin{figure}[!t]
	\centering
	\includegraphics[scale=0.62]{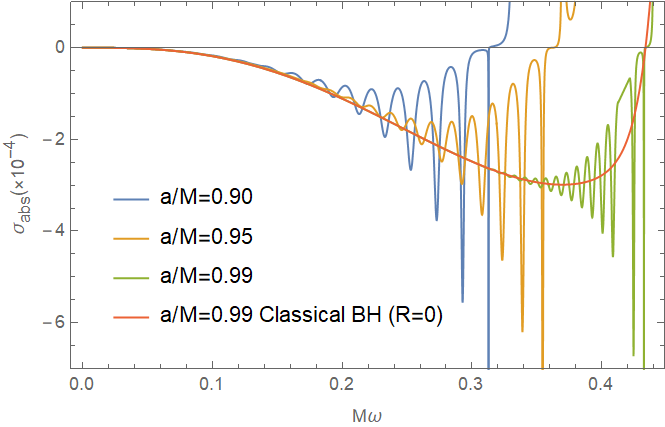}
	\\
	\includegraphics[scale=0.62]{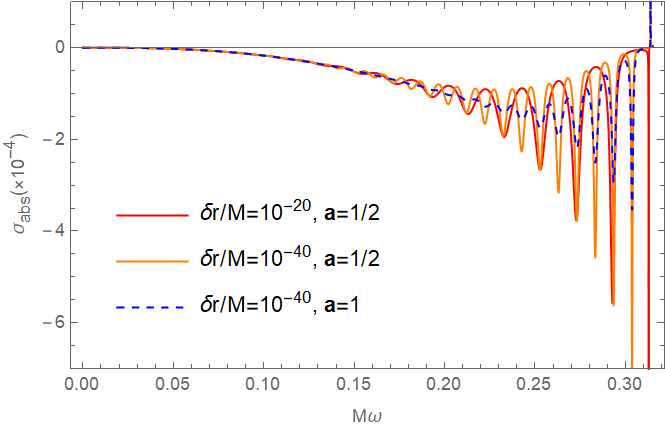}
	\caption{Absorption cross-section as a function of $M\omega$ with variation of $a/M$, $\delta r/M$, and \textbf{a}. For top plot we set $\textbf{a}=1/2$ and $\delta r/M=10^{-20}$. For bottom plot we set $a/M=0.99$. We use $m=l=1$ for both plots.}
	\label{fig:pabs}
\end{figure}

In the CFT description, we again consider CFT living on a circle with finite length $L$ and look into the finite size effects on the boundary. The absorption cross-section can be defined using Fermi's golden rule given by
\begin{align}
	\sigma_{abs} \sim \int &dt_Ldt_R e^{-i\omega_Lt_L}e^{-i\omega_Rt_R} (G(t_L-i\epsilon,t_R-i\epsilon)\nonumber\\
	&-G(t-L+i\epsilon,t_R+i\epsilon))
\end{align}
The $\pm i\epsilon$ determine the poles that contribute while performing the integration. Using the same way to perform the integration in (\ref{2ptftorus}), we obtain \cite{DeyAfshordiPRD2020}
\begin{align}\label{abscft}
	\sigma_{abs} \sim & \omega^{2l-1}T_L^{2h_L-1}T_R^{2h_R-1}\sinh\left(\frac{\tilde{\omega}_L}{2T_L}+\frac{\tilde{\omega}_R}{2T_R}\right)\nonumber\\
	&\times \left|\Gamma\left(h_L+i\frac{\tilde{\omega}_L}{2\pi T_L}\right)\Gamma\left(h_R+i\frac{\tilde{\omega}_R}{2\pi T_R}\right)\right|^2\nonumber\\ &\times\frac{1-e^{-2\textbf{a}\left|\frac{\tilde{\omega}_L}{T_L}+\frac{\tilde{\omega}_R}{T_R}\right|}}{\left|1-e^{i2\pi L(\tilde{\omega}_L+\tilde{\omega}_R)-\textbf{a}\left|\frac{\tilde{\omega}_L}{T_L}+\frac{\tilde{\omega}_R}{T_R}\right|}\right|^2}
\end{align}
This result agrees with absorption cross-section from gravity calculation (\ref{absgrav}) , if we choose
\begin{align}\label{identify}
	h_L &= h_R = \frac{1}{2}+\beta, & T_L &= \frac{1}{2\pi}, & T_R &= \frac{\tau_H}{4\pi}, \nonumber
\end{align}
\begin{align}
	\omega_L &= m, & \omega_R = \tilde{\omega}-m,
\end{align}
along with (\ref{toruslength}). The CFT temperatures and frequencies consistent with (\ref{cfttemp}) and (\ref{cftfrequency}) in near-extremal, near-superradiant limit that we have defined earlier. From Fig. (\ref{fig:pabs}), the absorption cross-section is negative in low-frequency because the superradiant condition  is similar to classical black hole.The supperradiance amplification is higher when $a/M\to1$. On the other hand, the unique feature of ECO's cross-section is the oscillation that comes from the exponential factor of (\ref{abscft}). This feature corresponds to resonances at QNM frequencies. Similar to QNMs, the oscillation also depends on $\delta r/M$, $a/M$, and \textbf{a}. These parameters influence amplitudes and phases of the oscillation. Specifically, the membrane position $\delta r/M$ determines the phase while the reflectivity (in this case depends on \textbf{a} and $a/M$) determines the amplitudes. If the exponential factor vanishes, the cross-section will reduce to that of classical black holes in near-extremal condition like in Ref.\cite{Bredberg2010}. We can also see that when $a/M\to1$, the oscillatory feature starts to disappear in lower frequencies. Thus, we can say that in low-frequency limit for $a\simeq M$ the Boltzmann reflectivity is suppressed by the spin of the ECO.

\section{Higher spin perturbations}
\label{sec6}
In the previous sections, we have shown that the absorption cross-section and QNM spectrums of near-extremal Kerr-like ECOs for scalar perturbation can be reproduced by the CFT. However, we can extend our discussion to general spin particle's perturbation such as electromagnetic and gravitation. To explore these perturbation or non-vanishing spin in general, we can apply the Newman-Penrose (NP) formalism \cite{Newman1962}. The NP null tetrad of Kerr $e_a^\mu=(l^\mu,n^\nu,m^\mu,m^{*\mu})$, in the basis $(\hat{t},\hat{r},\theta,\hat{\phi})$ is 
\begin{align}
	&l^\mu = \left(\frac{\hat{r}^2+a^2}{\Delta},1,0,\frac{a}{\Delta}\right),\nonumber\\
	&n^\mu = \frac{1}{2(\hat{r}^2 + a^2\cos^2\theta)}(\hat{r}+a,-\Delta,0,a),\\
	&m^\mu = \frac{1}{\sqrt{2}(\hat{r}+ia\cos\theta)}\left(ia\sin\theta,0,1,\frac{i}{\sin\theta}\right),\nonumber
\end{align}
with non-vanishing inner products
\begin{equation}
	l.n= -m.m^*=-1.
\end{equation}

The equation of motion of the perturbations in Kerr spacetime as its background are described in terms of Teukolsky's master equations \cite{Teukolsky1973a}. it turns out that this equation is separable for angular and radial functions. The wave function can be decomposed into the form
\begin{equation}
	\psi_s = e^{-i\omega\hat{t}+im\hat{\phi}}R_s(\hat{r})S_s(\theta)
\end{equation}
where $s$ is "spin-weight" of the field, $\psi_s$ are related to electromagnetic field strength and Weyl tensor for spin-1 $(|s|=1)$ and spin-2 $(|s|=2)$ perturbations. The angular and radial functions satisfy,
\begin{align}
	\frac{1}{\sin\theta}\frac{d}{d\theta}&\left(\sin\theta\frac{d}{d\theta}S_s\right)	+\left(\Lambda_{lms}-a^2\omega^2\sin^2\theta-\frac{m^2}{\sin^2\theta}\right.\nonumber\\
	&\left.-2a\omega s\cos\theta-\frac{s^2+2ms\cos\theta}{\sin^2\theta}\right)S_s=0,
\end{align}
and
\begin{align}
	\Delta^{-s}\frac{d}{d\hat{r}}&\left(\Delta^{s+1}\frac{d}{d\hat{r}}R_s\right)\nonumber\\
	&+\left(\frac{H^2-2is(\hat{r}-M)H}{\Delta}+4is\omega\hat{r}-\lambda\right)R_s=0,
\end{align}
where $\Lambda_{lms}$ is a separation constant, $\lambda=\Lambda_{lms}-2am\omega-s(s+1)$, and $H=\omega(\hat{r}^2+a^2)-am $.

In near-extremal condition, the radial equation in terms of $x$ and $\tau_H$ is
\begin{equation}\label{radialgenspin}
	x(x+\tau_H){R_s}''+(s+1)(2x+\tau_H){R_s}'+V_sR_s=0,
\end{equation}
where
\begin{eqnarray}
	V_s &= &\frac{(mx^2+2mx+\tilde{\omega}\tau_H)^2}{4x(x+\tau_H)}+2ims(1+x)-\lambda\nonumber\\
	&-&\frac{is(2x+\tau_H)(mx^2+2mx+\tilde{\omega}\tau_H)}{4x(x+\tau_H)}.\
\end{eqnarray}

\subsection{Far region}
In the far region $x\gg\tau_H$, the radial equation is
\begin{equation}
	x^2{R_s}''+(s+1)2x{R_s}'+V_s^{far}R_s=0,
\end{equation}
where
\begin{equation}
	V_s^{far} = -\Lambda_{lms}+m^2+\frac{m^2}{4}(x+2)^2+imsx+s(s+1).
\end{equation}
The solution to above equation is
\begin{align}
	R_s^{far}=&A_se^{-i\frac{mx}{2}}x^{-\frac{1}{2}+\beta_s-s}\nonumber\\
	&\times M\left(\frac{1}{2}+\beta_s-s+im,1+2\beta_s,imx\right)\nonumber\\
	&+B_s(\beta_s\to-\beta_s),
\end{align}
where $M(a,b,z)$ is Kummer function and
\begin{equation}
	\beta_s^2 = \frac{1}{4}+\Lambda_{lms}-2m^2.
\end{equation}
Again, we will only focus the case of real $\beta_s$. Take the solution to the matching region, $M(a,b,z)\to1$, it becomes
\begin{equation}\label{radialsolfar1s}
	R_s^{far} \sim A_sx^{-\frac{1}{2}+\beta_s-s}+B_sx^{-\frac{1}{2}-\beta_s-s}.
\end{equation}

\subsection{Near region}
In the near region $x\ll1$, the radial equation is (\ref{radialgenspin}) with
\begin{align}
	V_s^{near} = &\frac{(2mx+\tilde{\omega}\tau_H)^2-is(2x+\tau_H)(2mx+\tilde{\omega}\tau_H)}{4x(x+\tau_H)}\nonumber\\
	&+2ims+m^2+s(s+1)-\Lambda_{lms}.
\end{align}
The solution in the matching region is
\begin{align}\label{radialsolnear1s}
	&R_s^{near}\sim \tau_H^{\frac{1}{2}+\beta_s}x^{-\frac{1}{2}-\beta_s-s}\Gamma(-2\beta_s)\nonumber\\
	&\times\left[C_s\frac{\Gamma\left(1-s-i\tilde{\omega}\right)}{\Gamma\left(\frac{1}{2}-\beta_s-s-im\right)\Gamma\left(\frac{1}{2}-\beta_s-i(\tilde{\omega}-m)\right)}\tau_H^{-i\frac{\tilde{\omega}}{2}}\right.\nonumber\\
	&+\left.D_s\frac{\Gamma\left(1+s+i\tilde{\omega}\right)}{\Gamma\left(\frac{1}{2}-\beta_s+s+im\right)\Gamma\left(\frac{1}{2}-\beta_s+i(\tilde{\omega}-m)\right)}\tau_H^{i\frac{\tilde{\omega}}{2}+s}\right]\nonumber\\
	&+(\beta_s\to-\beta_s).
\end{align}
\subsection{Matching region}
The boundary condition at the membrane can be defined as
\begin{equation}\label{membounds}
	\mathcal{R}e^{i\pi\delta}=\frac{D_s}{C_s}x_0^{i\tilde{\omega}}
\end{equation}
Comparing coefficient from (\ref{radialsolfar1s}) and (\ref{radialsolnear1s}), we get
\begin{align}
	\frac{A_s}{C_s} = &\tau_H^{\frac{1}{2}-\beta_s-i\frac{\tilde{\omega}}{2}}\Gamma(2\beta_s) \left[Q_{s-}+\left(\frac{x_0}{\tau_H}\right)^{-i\tilde{\omega}-s}\mathcal{R}e^{i\pi\delta}Q_{s+}\right]\\
	\frac{B_s}{C_s} = &\tau_H^{\frac{1}{2}+\beta_s-i\frac{\tilde{\omega}}{2}}\Gamma(-2\beta_s) \left[P_{s-}+\left(\frac{x_0}{\tau_H}\right)^{-i\tilde{\omega}-s}\mathcal{R}e^{i\pi\delta}P_{s+}\right]
\end{align}
where
\begin{align}
	Q_{s\pm} &= \frac{\Gamma\left(1\pm s\pm i\tilde{\omega}\right)}
	{\Gamma\left(\frac{1}{2}+\beta_s\pm s\pm im\right)\Gamma\left(\frac{1}{2}+\beta_s\pm i(\tilde{\omega}-m)\right)}\\
	P_{s\pm} &= \frac{\Gamma\left(1\pm s\pm i\tilde{\omega}\right)}
	{\Gamma\left(\frac{1}{2}-\beta_s\pm s\pm im\right)\Gamma\left(\frac{1}{2}-\beta_s\pm i(\tilde{\omega}-m)\right)}
\end{align}
\subsection{Absorption cross-section and QNMs}
Following the way in section \ref{sec2}, \textcolor{black}{with only outgoing waves in infinity we have
\begin{equation}\label{DCs}
	\frac{D_s}{C_s}=-\frac{P_{s-}+\sigma_s Q_{s-}}{P_{s+}+\sigma_s Q_{s+}},
\end{equation}
with
\begin{equation}
	\sigma_s = (\tau_H/im)^{-2\beta}\frac{\Gamma(2\beta_s)\Gamma(1+2\beta_s)\Gamma(1-\beta_s-s-im)}{\Gamma(-2\beta_s)\Gamma(1-2\beta_s)\Gamma(1+\beta_s-s-im)}.
\end{equation}
Thus we get
\begin{equation}
	\mathcal{R}e^{i\pi\delta}=-\frac{P_{s-}+\sigma_s Q_{s-}}{P_{s+}+\sigma_s Q_{s+}}x_0^{i\tilde{\omega}}.
\end{equation}
The QNM spectrum in the low-frequency limit is}
\begin{align}\label{qnms}
	\omega-m\Omega_H \simeq &\frac{1}{2r_0^*}\pi\left(2n+\frac{(-1)^s+1}{2}+\delta\right)\nonumber\\
	&\times\left(1-\frac{i\times \text{sgn}[2n+\frac{(-1)^s+1}{2}+\delta]}{4r_0^*\gamma T_H}\right).
\end{align}
This result is consistent with CFT when we define
\begin{align}\label{deltas}
	\delta=-\frac{(-1)^s+1}{2}-2mL
\end{align}
and $L,\textbf{a}$ are the same as given in Eq. (\ref{toruslength}) and (\ref{agamma}). On the other hand, the absorption cross-section is
\begin{align}
	\sigma_{abs} &\sim \frac{(\tau_H)^{2\beta_s}}{\pi\Gamma(2\beta_s)^2}\sinh\left(\pi \tilde{\omega}\right)\nonumber\\
	&\times\left|\Gamma\left(\frac{1}{2}+\beta_s-s+im\right)\Gamma\left(\frac{1}{2}+\beta_s+i(\tilde{\omega}-m)\right)\right|^2\nonumber\\
	&\times\frac{1-|\mathcal{R}|^2}{\left|1-\mathcal{R}e^{-2ir_0^*(\omega-m\Omega_H)+i\pi\delta}\right|^2}
\end{align}
This result agrees with the CFT result when we choose
\begin{align}
	h_R &= \frac{1}{2}+\beta_s, & h_L &= h_R-s,
\end{align}
and $T_L,T_R,\omega_L,\omega_R$ are also given by Eq. (\ref{identify}). We can take $s=0$ and this will reduce to Eq. (\ref{qnm}) and Eq. (\ref{absgrav}) for scalar perturbation. Furthermore, we can find the results for electromagnetic and gravitational perturbation if we take $s=1$ and $s=2$, respectively.

\begin{figure*}[!t]
	\centering
	\includegraphics[scale=0.68]{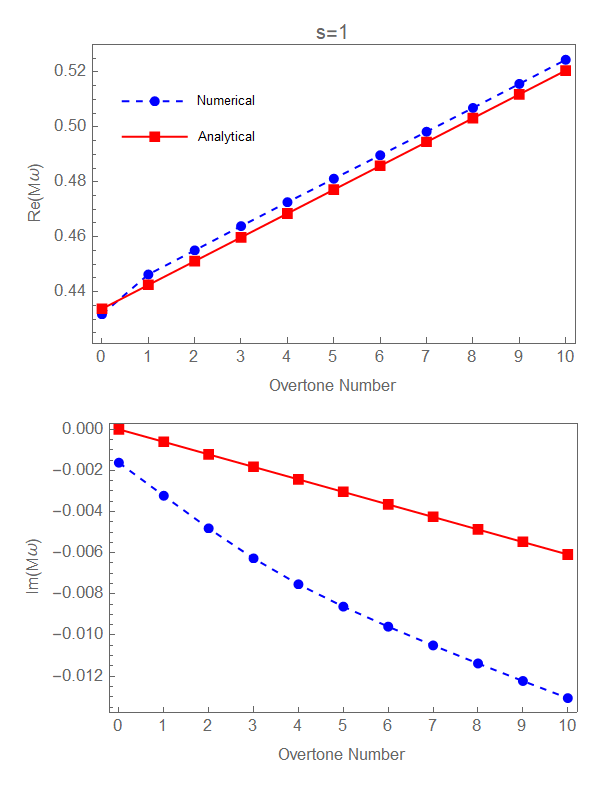}
	\hspace{1em}
	\includegraphics[scale=0.68]{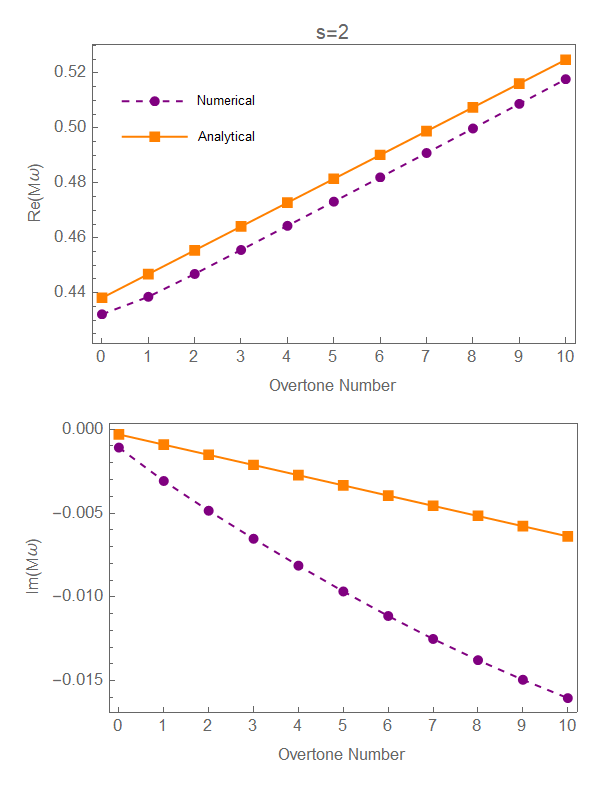}
	\caption{\textcolor{black}{Real and Imaginary part of QNM from numerical and analytical methods for $s=1$ (left) and $s=2$ (right). For both panels we set $a/M=0.99$, $\delta r/M=10^{-20}$, $m=1$, $\delta=0$, and $\textbf{a}=1/2$.}}
	\label{fig:qnms}
\end{figure*}

Each perturbation has different observational signature that we can see from its QNM and absorption cross-section. \textcolor{black}{In Fig. (\ref{fig:qnms}) we compare QNM from numerical and analytical methods for $s=1$ and $s=2$ with $a/M=0.99$ and $\delta r/M=10^{-20}$. We can see that analytical method with low-frequency approximation is more plausible for the real part. Meanwhile, because the approximation is applied to simplify Eq. (\ref{DCs}) which mostly contribute to the imaginary part, we can see that analytical result of it is not reliable for higher overtones in this case. The analytical approximation might not be valid due to the fact that the QNM in near-extremal limit and high overtone is not in low-frequency regime.} Other than that, we still get $Im(\omega)\ll Re(\omega)$ which will lead into echoes production. Furthermore, for all $s$, the imaginary part of QNM is negative which means the instability is avoided.

\begin{figure}[!t]
	\centering
	\includegraphics[scale=0.65]{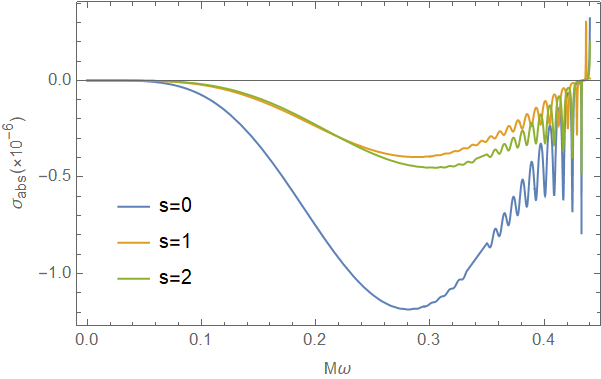}
	\caption{Absorption cross-section as a function of $M\omega$ for various $s$. We set $a/M=0.99$, $m=1$, $l=2$, $\gamma=1$, and $\delta r/M=10^{-20}$.}
	\label{fig:pabss}
\end{figure}

In Fig. (\ref{fig:pabss}), we can see that gravitational perturbation has more negative $\sigma_{abs}$ than electromagnetic perturbation, but scalar perturbation has the most negative $\sigma_{abs}$ where usually larger spin is the one that more affected by superradiant scattering \cite{Teukolsky1973b,Teukolsky1974}. This is because we set $m=1$ and $l=2$ for all spin-$s$ so the amplification from superradiant is not maximum.  Normally to obtain the maximum effect of superradiant scattering, we need to set $l=m$ such as $l=m=1$ for scalar and electromagnetic and $l=m=2$ for gravitational waves \cite{Brito2015}. However, in this work we can only consider real $\beta_s$ to match the results of absorption cross-section with CFT. Since $\beta_s$ depends on $l$, $m$, and $s$ in the separation constant $\Lambda_{lms}$, most of $l=m$ combination for electromagnetic and gravitation perturbation is not compatible for our calculation. Because of this reason, we cannot show cases where the electromagnetic and gravitational waves are amplified more than the scalar one. Furthermore, when we matching the gravity result with CFT, we only consider cross-section equal up to a factor independent of frequencies \cite{ChenChu2010}. Other normalization factor might depends on spin-$s$ and impact the superradiance. We also notice that in the electromagnetic case, the oscillation has different phase $\pi$ compare to the scalar and gravitational case which corresponds to $\delta$ in Eq. (\ref{deltas}).

\section{Echo Time-delay}
\label{sec7}

\begin{figure}[!t]
	\centering
	\includegraphics[scale=0.55]{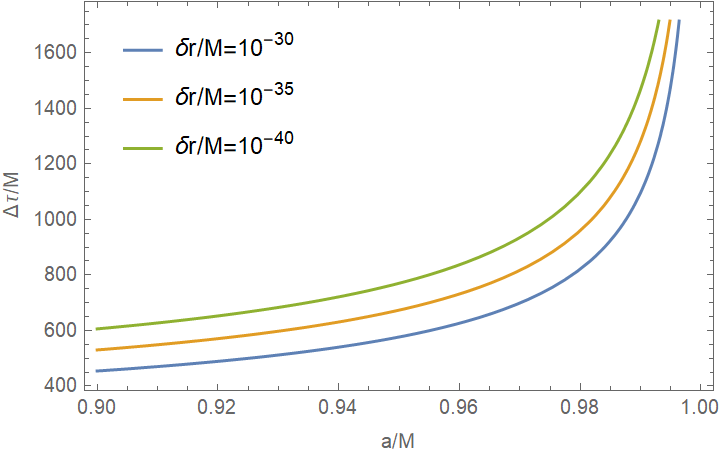}
	\caption{\textcolor{black}{Echo time-delay as a function of $a/M$ near extremality with variation of $\delta r/M$.}}
	\label{fig:deltatau}
\end{figure}

\begin{figure}[!t]
	\centering
	\includegraphics[scale=0.6]{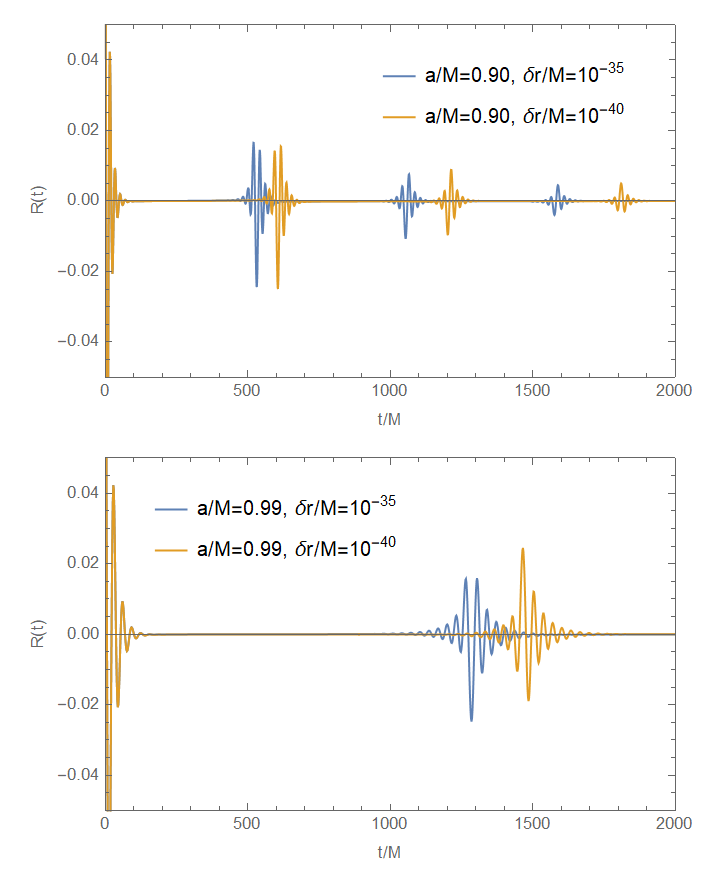}
	\caption{\textcolor{black}{Gravitational echoes waveform in time domain. We set $a/M=0.90$ for top plot and $a/M=0.99$ for bottom plot.}}
	\label{fig:echotest}
\end{figure}

Since there is a reflective membrane, the ingoing wave is partially reflected and trapped between the membrane and angular momentum barrier. Eventually the waves will tunnel through the barrier as repeating echoes. The time gap between two consecutive echoes is time for the wave to travel from the angular momentum barrier to the membrane and travel back again. This time gap or time-delay can be stated in terms of the ECO parameters and the position of the reflective membrane. The echoes time-delay can be defined as
\begin{equation}\label{techodef}
	\Delta\tau=2|r^*_0|
\end{equation}
This expression indicates that $\Delta\tau$ is sensitive to the position of the membrane. In terms of distance of the membrane from the usual position of horizon $\delta r=\hat{r}_0-r_+$, Eq. (\ref{techodef}) becomes
\begin{equation}\label{techodef2}
	\Delta\tau \simeq \pm\frac{4Mr_+}{(r_+-r_-)} \ln\left(\frac{\delta r}{r_+-r_-}\right),
\end{equation}
where plus sign is for $\delta r > r_+-r_-$ and minus sign is for $\delta r < r_+-r_-$. However, we only consider the later case because as we see in Fig. (\ref{fig:deltatau}) and Fig. (\ref{fig:echotest}) larger $\delta r$ should have shorter time-delay, since the distance between the angular momentum barrier and the reflective membrane is shorter. We get a similar result to non-extreme case \cite{DeyAfshordiPRD2020} with slight difference on the logarithmic factor. This is caused by different coordinate transformation when converting the position of the membrane in tortoise coordinate to the proper distance, where in this case we use near-extremal limit (\ref{nearexcoord}). Nonetheless, we still attain the sensitivity of the echo time-delay to the position of the membrane. In addition to that, it also sensitive to the extremality of the ECOs. 

The length of the torus in terms of echo time-delay is
\begin{equation}
	L=\frac{1}{4\pi M}\frac{\Delta\tau}{2}=-\frac{r_+}{2\pi(r_+-r_-)} \ln\left(\frac{\delta r}{r_+-r_-}\right).
\end{equation}
Thus, we obtain
\begin{equation}\label{deltar}
	\delta r=(r_+-r_-)e^{-\frac{2\pi(r_+-r_-)}{r_+}L}.
\end{equation}
This relation shows that the position of the membrane from the horizon associated with the size of the torus. When $\delta r\to0$, we can have $L\to\infty$ which corresponds to classical case. If we expand (\ref{deltar}) and take up to the first order, we have relation between $\delta r$ and dimensionless Hawking temperature
\begin{equation}\label{deltartauh}
	\delta r \simeq (r_+-r_-)\left(1-\frac{2\pi(r_+-r_-)}{r_+}L\right)\simeq\tau_Hr_+.
\end{equation}
Since $\delta r$ is assumed in order of Planck length for stability, this relation fits with near-extremal condition $\tau_H\ll1$.

In Fig. (\ref{fig:deltatau}) we show the plot of echo time-delay as a function of $a/M$ near extremality with variation of $\delta r/M$. We can see that when $a/M\to1$, echo time-delay will approach infinity, in other words there are no echoes. This can be seen as for extremal case, the reflectivity is suppressed by the spin of the ECO thus ECO absorbs the incoming wave like classical black hole and does not produce echoes. Moreover, echo time-delay does not depend on the particle spin, $s$. Since echo time-delay is defined as how long it takes for a wave to travel across the ergoregion twice. For massless scalars, photons, and gravitons that travel with approximately same speed, these will have same time-delay. 

To see explicitly the existence of echoes, we will show in this work the waveform of gravitational echoes in time domain. Echoes from ECO is constructed by reprocessing the standard black holes post-merger ringdown to include information about the physical properties of the modified horizon \cite{Testa2018}, such as the membrane reflectivity. We adopt model from \cite{Magio2019b} for the analytical model of gravitational-wave echoes from spinning ECO, in frequency domain the ringdown + echo waveform is given by
\begin{equation}
	\tilde{Z}(\omega)=\tilde{Z}_{BH}+\frac{\mathcal{R}_{BH}\mathcal{R}e^{-2ir_0^*(\omega-m\Omega_H)+i\pi\delta}}{1-\mathcal{R}_{BH}\mathcal{R}e^{-2ir_0^*(\omega-m\Omega_H)+i\pi\delta}}\tilde{Z}_{BH}.
\end{equation}	
The first term is the single mode black hole ringdown, where $\tilde{Z}_{BH}(\omega)$ is the responses of Kerr black hole at infinity to a source. While the second term is the echo, where $\mathcal{R}_{BH}(\omega)$ is the black hole reflection coefficient of angular momentum barrier. The time domain waveform is computed through inverse Fourier transform,
\begin{equation}
	R(t)=\frac{1}{\sqrt{2\pi}}\int_{-\infty}^{\infty}d\omega \tilde{Z}(\omega)e^{-i\omega t}.
\end{equation}
\textcolor{black}{In Fig. (\ref{fig:echotest}), we can see that for each variation the gap between each echoes is the same, approximately (\ref{techodef2}).} The variation of $a/M$ and $\delta r/M$ in the time-delay shows that the echoes is sensitive to ECO's spin and membrane's location. We can see more clearly that for larger $\delta r/M$ and for smaller $a/M$ the time-delay is shorter. In late times, low-frequency modes is dominant and become the source of the echoes as the black holes ringdown is already decayed.

\section{Conclusions}
\label{conc}

In this paper, we analyzed the CFT dual of Kerr-like ECO in near-extremal condition. Due to quantum gravitational effects, the horizon is replaced by partially reflective membrane that placed slightly outside the would-be horizon. To keep the reflectivity in the near-extremal limit non-zero, we only considered fields with energies near-superradiant bound. We imposed this reflective boundary conditions for scalar field perturbation. The QNM spectrum and absorption cross-section is then obtained by solving the propagation of massless scalar field modes. We compared the results by using CFT dual computation as has been done for generic Kerr ECO. The CFT calculation of modified near-horizon region is done by considering dual field theory that lives on a finite toroidal two-manifold, where we kept the spatial and temporal periodicity finite. This method of modification can be interpreted as finite size/finite $N$ effects in AdS/CFT.

We showed the consistency between the QNMs and the absorption cross-section computed from the gravity side and similar quantities computed from the dual CFT. The QNMs in near-extremal condition is in line with non-extreme case, where the differences are in the value of torus length $L$ and phase shift $\delta$. We reproduced Boltzmann reflectivity from imaginary parts of QNMs from CFT, which agreed with non-extreme case. The imaginary part of QNMs also related with instability of ECO. The instability in near-extremal regime is suppressed by the partially reflective membrane as long as we choose the modular parameter \textbf{a} to be positive. Besides QNMs, the absorption cross-section possessed oscillatory features that start to disappear when $a/M\to1$. This showed that the reflectivity of ECO is suppressed by the spin of the ECO near extremality. The CFT side reproduced this features by choosing CFT quantities that is consistent with near-extremal condition. 

We also extended the calculation for higher spin fields such as photon and graviton. We solved the Teukolsky's master equation for general spin and imposed reflective boundary condition, then we obtained the QNMs and absorption cross-section. The particle spin contributes to the phase shift of the reflected waves. In the electromagnetic perturbation, the frequency has different phase $\pi$ compare to the scalar and gravitational perturbation. Moreover, in terms of CFT quantity the particle spin appears in the conformal weight. 

\textcolor{black}{The comparison for the QNMs from analytical and numerical methods has also been provided. It shows that the analytical approximation in low-frequency limit for scalar mode for the real and imaginary parts works plausibly. However, the analytical approximation for QNMs for higher spin fields does not work well as for the scalar mode. The analytical approximation might not be valid for higher spins due to the fact that the QNM in near-extremal limit and high overtone is not in low-frequency regime.}

In the end, we computed the echo time-delay produced by the propagation of the fields on the near-extremal Kerr ECO background. Echo time-delay is necessary observable from the gravitational echoes observation in the postmerger ringdown signal. The modification of near-horizon region could also be manifested in this observable. We showed that the echoes time-delay is sensitive to the position of the membrane and the extremality of the ECOs. It is very important to investigate the quantum corrections to the near-horizon geometry of near-extremal Kerr from the CFT side. It may lead to a better understanding of the microscopic theory. We believe that our calculation on the QNMs, absorption cross-section, and echo time-delay in near-extremal condition can be used to advance the study of ECO, especially with the relation with experiment since in the near future more experiments will run to improve the study of astrophysical objects. Furthermore, extending our analysis for higher spin fields provide the possibility of observing these physical properties of ECO such as through gravitational waves.

\begin{acknowledgments}
We would like to thank the members of Theoretical Physics Groups of Institut Teknologi Bandung for the discussion. M. Z. Djogama is supported by LPDP, Ministry of the Finance Republic of Indonesia. M. F. A. R. S. is supported by the Second Century Fund (C2F), Chulalongkorn University, Thailand. F. P. Zen is supported by financial funds of Ministry of Education, Culture, Research and Technology Republic of Indonesia.
\end{acknowledgments}

\appendix

\section{Scalar Wave Equation}
\label{app:scalwave}

In the far region $x\gg\tau_H$, the radial equation (\ref{radialeqkerr2}) becomes
\begin{equation}\label{radialeqfar}
	\partial_z^2R+\left(-\frac{1}{4}+\frac{(-im)}{z}+\frac{\left(\frac{1}{4}-\beta^2\right)}{z^2}\right)R=0,
\end{equation}
where we define $z=imx$ and
\begin{equation}
	\beta^2=\frac{1}{4}+K_l-2m^2.
\end{equation}
The solution for above equation is
\begin{eqnarray}
	R_{far}&=&Ae^{-\frac{1}{2}z}x^{-\frac{1}{2}+\beta}M\left(\frac{1}{2}+\beta+im,1+2\beta,z\right)\nonumber\\
	&&+B(\beta\to-\beta),
\end{eqnarray}
where $M(a,b,z)$ is Kummer function. In the matching region, the solution reads
\begin{equation}\label{radialsolfar1}
	R_{far} \sim Ax^{-\frac{1}{2}+\beta}+Bx^{-\frac{1}{2}-\beta}.
\end{equation}
While in the asymptotic region $x\gg1$,
\begin{equation}\label{farflat} 
	R_{far}\sim Z_{out}e^{\frac{1}{2}imx}x^{-1+im}+Z_{in}e^{-\frac{1}{2}imx}x^{-1-im},
\end{equation}
where
\begin{align}
	Z_{in} &= AC_++BC_-, \\
	Z_{out} &= A\tilde{C}_++B\tilde{C}_-, \nonumber\\
	C_\pm &= \frac{\Gamma(1\pm2\beta)}{\Gamma(\frac{1}{2}\pm\beta-im)}(-im)^{\frac{1}{2}\mp\beta-im}, \nonumber\\
	\tilde{C}_\pm &= \frac{\Gamma(1\pm2\beta)}{\Gamma(\frac{1}{2}\pm\beta+im)}(im)^{\frac{1}{2}\mp\beta+im}.\nonumber
\end{align}

In the near region $x\ll1$, the radial equation becomes
\begin{equation}
	x(x+\tau_H)R''+(2x+\tau_H)R'+V_l^{near}R=0,
\end{equation}
where
\begin{equation}
	V_l^{near}=\frac{(2mx+\tau_H\tilde{\omega})^2}{4x(x+\tau_H)}+m^2-K_l.
\end{equation}
The solution in this region can be written in form of hypergeometric functions $F(a,b,c;z)$, 
\begin{eqnarray}\label{radialsolnear}
	R_{near} &=&  Cx^{-\frac{i}{2}n}\left(\frac{x}{\tau_H}+1\right)^{i(\frac{n}{2}-m)}F\left(a,b, c;z\right)\nonumber\\
	& +&Dx^{\frac{i}{2}n}\left(\frac{x}{\tau_H}+1\right)^{i(\frac{n}{2}-m)}\nonumber\\
	&&\times F\left(a-c+1, b-c+1, 2-c; z\right),
\end{eqnarray}
where $a=1/2+\beta-im,$ $b=1/2-\beta-im,$ $c=1-in,$ and $z=-x/\tau_H$. In the matching region, the large-$x$ behavior of the solution is
\begin{align}\label{radialsolnear1}
	R_{near} &\sim \tau_H^{\frac{1}{2}+\beta}x^{-\frac{1}{2}-\beta}\Gamma(-2\beta)\nonumber\\
	&\times \left[C\frac{\Gamma\left(1-i\tilde{\omega}\right)}{\Gamma\left(\frac{1}{2}-\beta-im\right)\Gamma\left(\frac{1}{2}-\beta-i(\tilde{\omega}-m)\right)}\tau_H^{-\frac{i}{2}\tilde{\omega}}\right.\nonumber\\
	&\left.+D\frac{\Gamma\left(1+i\tilde{\omega}\right)}{\Gamma\left(\frac{1}{2}-\beta+im\right)\Gamma\left(\frac{1}{2}-\beta+i(\tilde{\omega}-m)\right)}\tau_H^{\frac{i}{2}\tilde{\omega}}\right]\nonumber\\
	&+(\beta\to-\beta).
\end{align}
While near the membrane $x\to0$
\begin{align}\label{radialsolnear2}
	R_{near} &\sim Cx^{-\frac{i}{2}\tilde{\omega}}+Dx^{\frac{i}{2}\tilde{\omega}}\nonumber\\
	&\sim Ce^{-i(\omega-m\Omega_H)r^*}+De^{i(\omega-m\Omega_H)r^*},
\end{align}
The first part on the right hand side of Eq. (\ref{radialsolnear2}) can be seen as the ingoing wave and the second part as the outgoing wave.

The new boundary condition at the membrane can be defined in terms of asymptotic amplitudes as
\begin{equation}\label{membound}
	\mathcal{R}e^{i\pi\delta}=\frac{D}{C}x_0^{i\tilde{\omega}},
\end{equation}
where $x_0=x(\hat{r}_0)$ is the position of the membrane and $\delta$ is a phase shift determined by the properties of the ECO. In the matching region $\tau_H\ll x\ll 1$, solutions from far and near regions are both valid. By comparing the amplitudes from (\ref{radialsolfar1}) and (\ref{radialsolnear1}), we obtain
\begin{align}
	&\frac{A}{C} = \tau_H^{\frac{1}{2}-\beta-\frac{i}{2}\tilde{\omega}}\Gamma(2\beta)\label{ACmatching} \left[Q_-+\left(\frac{x_0}{\tau_H}\right)^{-i\tilde{\omega}}\mathcal{R}e^{i\pi\delta}Q_+\right],\\
	&\frac{B}{C} = \tau_H^{\frac{1}{2}+\beta-\frac{i}{2}\tilde{\omega}}\Gamma(-2\beta) \left[P_-+\left(\frac{x_0}{\tau_H}\right)^{-i\tilde{\omega}}\mathcal{R}e^{i\pi\delta}P_+\right],
\end{align}
where
\begin{align}
	&P_\pm = \frac{\Gamma\left(1\pm i\tilde{\omega}\right)} {\Gamma\left(\frac{1}{2}-\beta\pm im\right)\Gamma\left(\frac{1}{2}-\beta\pm i(\tilde{\omega}-m)\right)},\\
	&Q_\pm = \frac{\Gamma\left(1\pm i\tilde{\omega}\right)} {\Gamma\left(\frac{1}{2}+\beta\pm im\right)\Gamma\left(\frac{1}{2}+\beta\pm i(\tilde{\omega}-m)\right)}.
\end{align}

\section{CFT two-point function on the torus}
\label{app:2pttorus}

For the two-point function on a Euclidean plane with coordinates $x_1$ and $x_2$, we can define complex coordinates such as
\begin{equation}
	w=x_1+ix_2, ~~~ \bar{w}=x_1-ix_2.
\end{equation}
Primary field $\phi(w,\bar{w})$ with conformal dimension $(h,\bar{h})$ under conformal transformation $w\to f(t_R), \bar{w}\to \bar{f}(t_L)$ becomes
\begin{equation}
	\phi(t_R,t_L)=\left|\frac{df'(w)}{dw}\right|^h \left|\frac{d\bar{f}'(\bar{w})}{d\bar{w}}\right|^{\bar{h}} \phi(f(w),\bar{f}(\bar{w})).
\end{equation}
Based on the transformation properties of local CFT operator, the two-point function will also transform similarly under conformal transformation. Thus we can have fix form of CFT two-point function of operator $\mathcal{O}(w,\bar{w})$ as
\begin{equation}
	\langle\mathcal{O}_1(w_1,\bar{w}_1)\mathcal{O}_2(w_2,\bar{w}_2)\rangle=\frac{C_{12}}{(w_1-w_2)^{2h_R}(\bar{w}_1-\bar{w}_2)^{2h_L}},
\end{equation}
where $C_{12}$ is constant and $(h_R,h_L)$ is conformal dimension of $\mathcal{O}_{1,2}$.

Now for a two-dimensional Euclidean torus, we can define it with coordinates $(x,\tau_E)$ and periodicity $(x,\tau_E)\to(x+L,\tau_E+T^{-1})$. The torus is characterized by modular parameter $\tau=iT^{-1}/L$ and also complex holomorphic (antiholomorphic) coordinate $t_R=x+i\tau_E$ $(t_L=x-i\tau_E)$ with periodicity $t_R\to t_R+nL+imT^{-1}$ $(t_L\to t_R+nL-imT^{-1})$, for $n,m\in\mathbb{Z}$. These characteristic is invariant under modular transformation such as
\begin{equation}
	\tau'=\frac{\textbf{a}\tau+\textbf{b}}{\textbf{c}\tau+\textbf{d}}, ~~~ t_R'=\frac{t_R}{\textbf{c}\tau+\textbf{d}}, ~~~ \textbf{a}\textbf{d}-\textbf{b}\textbf{c}=1.
\end{equation}
The relation between $w$ and $t_R$ can be defined as $w=e^{it_R/(\textbf{c}\tau+\textbf{d})}$. In terms of torus coordinate, the two-point function can be written as
\begin{align}
	\langle\mathcal{O}_1&(t_R,t_L)\mathcal{O}_2(0,0)\rangle_{SL(2,\mathbb{Z})}\nonumber\\
	&=\sum_{n\in\mathbb{Z}} \left|2(\textbf{c}\tau+\textbf{d})\sin\pi\left[\frac{t_R+nL(\textbf{a}\tau+\textbf{b})}{L(\textbf{c}\tau+\textbf{d})}\right]\right|^{-2h_R}\nonumber\\
	&\times \left|2(\textbf{c}\bar{\tau}+\textbf{d})\sin\pi\left[\frac{t_L+nL(\textbf{a}\bar{\tau}+\textbf{b})}{L(\textbf{c}\bar{\tau}+\textbf{d})}\right]\right|^{-2h_L},
\end{align}
where $\tau=iT_R^{-1}/L, \bar{\tau}=iT_L^{_1}/L$ and $h_R,h_L$ is conformal weights.

We want to compute the two-point function for a specific set values of parameters $(\textbf{a},\textbf{b},\textbf{c},\textbf{d})$, which correspond to a specific geometric configuration. For that we perform analytical continuation $\tau_E\to it$, thus $t_R=x-t$ and $t_L=x+t$. The two-point function now is given as
\begin{align}\label{2pttorus}
	\langle\mathcal{O}_1&(t_R,t_L)\mathcal{O}_2(0,0)\rangle_{SL(2,\mathbb{Z})}\nonumber\\
	&=\sum_{n\in\mathbb{Z}} \left[\frac{L^2}{\textbf{c}^2T_R^{-2}+\textbf{d}^2L^2}\right]^{2h_R}\left[\frac{L^2}{\textbf{c}^2T_L^{-2}+\textbf{d}^2L^2}\right]^{2h_L}\nonumber\\
	&\times \frac{(\pi/L)^{2h_R}(\pi/L)^{2h_L}}{[\sinh\pi(x_R)]^{2h_R}[\sinh\pi(x_L)]^{2h_L}},
\end{align}
where
\begin{align}
	x_{R,L}=& \frac{T_{R,L}^{-1}}{\textbf{c}^2T_{R,L}^{-2}+\textbf{d}^2L^2} (\textbf{c}t_{R,L}\pm nL)\nonumber\\
	&-i\frac{L}{\textbf{c}^2T_{R,L}^{-2}+\textbf{d}^2L^2} \nonumber\\
	&\times\left(\textbf{d}t_{R,L}+n\frac{\textbf{a}\textbf{c}T_{R,L}^{-2}+\textbf{b}\textbf{d}L^2}{L}\right).
\end{align}

\end{document}